# The Great Space Weather Event during February 1872 Recorded in East Asia

Hisashi Hayakawa* (1-2), Yusuke Ebihara (3-4), David M. Willis (2, 5), Kentaro Hattori (6), Alessandra S. Giunta (2), Matthew N. Wild (2), Satoshi Hayakawa (7), Shin Toriumi (8), Yasuyuki Mitsuma (9), Lee T. Macdonald (10), Kazunari Shibata (11), and Sam M. Silverman (12)

(1) Graduate School of Letters, Osaka University, 5600043, Toyonaka, Japan (JSPS Research Fellow).

(2) Science and Technology Facilities Council, RAL Space, Rutherford Appleton Laboratory, Harwell Campus, Didcot, OX11 0QX, UK

(3) Research Institute for Sustainable Humanosphere, Kyoto University, Uji, 6100011, Japan

(4) Unit of Synergetic Studies for Space, Kyoto University, Kyoto, 6068306, Japan

(5) Centre for Fusion, Space and Astrophysics, Department of Physics, University of Warwick, Coventry CV4 7AL, UK

(6) Graduate School of Science, Kyoto University, Kyoto, Kitashirakawa Oiwake-cho, Sakyo-ku, Kyoto, 6068502, Japan

(7) Faculty of Engineering, The University of Tokyo, 1130033, Tokyo, Japan

(8) National Astronomical Observatory of Japan, 1818588, Mitaka, Japan (NAOJ Fellow).

(9) Graduate School of Arts and Sciences, The University of Tokyo, 1538902, Tokyo, Japan

(10) Museum of the History of Science, Oxford OX1 3AZ, UK

(11) Kwasan Observatory, Kyoto University, 6078471, Kyoto, Japan

(12) 18 Ingleside Road, Lexington, MA 02420, USA

*email: hayakawa@kwasan.kyoto-u.ac.jp

**Abstract**

The study of historical great geomagnetic storms is crucial for assessing the possible risks to the technological infrastructure of a modern society, caused by extreme space–weather events. The normal benchmark has been the great geomagnetic storm of September 1859, the so-called 'Carrington Event'. However, there are numerous records of another great geomagnetic storm in February 1872. This storm, about 12 years after the Carrington Event, resulted in comparable magnetic disturbances and auroral displays over large areas of the Earth. We have revisited this great geomagnetic storm in terms of the auroral and sunspot records in the historical documents from East Asia. In particular, we have surveyed the auroral records from East Asia and estimated the equatorward boundary of the auroral oval to be near 24.3° invariant latitude (ILAT), on the basis that the aurora was seen near the zenith at Shanghai (20° magnetic latitude, MLAT). These results confirm that this geomagnetic storm of February 1872 was as extreme as the Carrington Event, at least in terms of the equatorward motion of the auroral oval. Indeed, our results support the interpretation of the simultaneous auroral observations made at Bombay (10° MLAT). The East Asian auroral records have indicated extreme brightness, suggesting unusual precipitation of high-intensity, low-energy electrons during this geomagnetic storm. We have compared the duration of the East Asian auroral displays with magnetic observations in Bombay and found that the auroral displays occurred in the initial phase, main phase, and early recovery phase of the magnetic storm.

## 1. Introduction

The results of recent studies provide a timely warning that great geomagnetic storms could be catastrophic for a modern technological society that is highly dependent on space-based systems for many of the essential services, such as communications, security monitoring, space weather forecasts, and is also reliant on an extensive electrical infrastructure (e.g. Baker et al., 2008; Hapgood, 2011; Pulkkinen et al., 2012, 2017; Ngwira et al., 2013; Oughton et al., 2016; Riley and Love, 2018; Love et al., 2018). Moreover, the study of historical extreme space–weather events is closely related to the goals of existing spacecraft missions, such as the Solar Terrestrial Relations Observatory (STEREO: Kaiser et al., 2008; Rouillard et al., 2009), the Geostationary Operational Environmental Satellite (GOES)[1], and the Deep Space Climate Laboratory (DSCOVR)[2]. Likewise, there is significant synergy between the investigation of historical extreme space–weather events and the goals of forthcoming spacecraft missions, such as the Parker Solar Probe (Fox et al. 2016), due for launch in the summer of 2018, and the Solar Orbiter (Müller et al., 2013), due for launch in 2020. The aim of all these missions is to obtain an improved understanding of the connection and coupling between the Sun and the heliosphere, with particular attention being directed towards the resulting effects on the Earth, including potentially catastrophic effects.

It is known that large sunspot groups may sometimes be associated with significant solar flares and CMEs that cause intense magnetic storms on the Earth (e.g. Vaquero and Vázquez, 2009; Odenwald, 2015; Usoskin, 2017; Riley et al., 2018; Lockwood et al., 2018). It is believed that the Carrington storm in September 1859 is one of the largest magnetic storms ever recorded (e.g. Mayaud, 1980; Tsurutani et al., 2003; Lakhina and Tsurutani, 2016). Carrington (1859) and Hodgson (1859) observed a white-light flare in a large sunspot on September 1, 1859 and the resultant CMEs caused great auroral displays, which were visible even in areas of magnetic latitude (MLAT) as low as 20°-23° (Kimball, 1960; Tsurutani et al., 2003; Cliver and Svalgaard, 2004; Green and Boardsen, 2006; Cliver and Dietrich, 2013; Hayakawa et al., 2016b; Moreno Cárdenas et al., 2016; Lakhina and Tsurutani, 2016). This magnetic storm produced intense magnetic disturbances over much of the Earth, which resulted in a maximum negative intensity of ~1600 nT at Colaba (Tsurutani et al., 2003; Nevanlinna, 2008; Ribeiro et al., 2011; Cliver and Dietrich, 2013; Kumar et al., 2015), and significant damage to the telegraph network (Boteler, 2006; Cliver and Dietrich, 2013).

---

[1] https://www.nasa.gov/content/goes
[2] http://www.nesdis.noaa.gov/DSCOVR/

Nevertheless, the Carrington storm is not the sole extreme magnetic storm recorded in recent human history. Chapman (1957a) listed some outstanding auroral displays at low MLAT including a major storm in 1872, during the coverage of magnetic observations from the mid-19th century. Large magnetic storms identified both by magnetic observations and auroral observations have attracted significant interest and important investigations (e.g. Silverman, 1995, 2006, 2008; Silverman and Cliver, 2001; Nevanlinna, 2004, 2006, 2008; Shiokawa et al., 2005; Vaquero et al., 2008; Cliver and Dietrich, 2013; Viljanen et al., 2014; Lefèvre et al., 2016; Lockwood et al., 2016, 2018; Vennerstrom et al., 2016; Love, 2017; Riley et al., 2018), while recent studies have identified extreme magnetic storms before magnetic observations became available continuously with eyewitness auroral records (e.g. Willis et al., 1996a, 2005, 2007; Willis and Stephenson, 2001; Hayakawa et al., 2017a, 2017b, 2017c, 2018; Ebihara et al., 2017).

It is noteworthy that another major storm occurred only one solar cycle after the Carrington storm of 1859. Cliver and Dietrich (2013) investigated historically outstanding magnetic storms and specifically considered large storms that occur once in 60 to 100 years. Curto et al. (2016) assumed the Carrington flare to be an X45 class flare and estimated its return period as 90 ± 60 years. The major storms of the 1770s associated with extreme auroral brightness, are also consistent in terms of their return period with the Carrington storm (Willis et al., 1996a; Hayakawa et al., 2017c, Ebihara et al., 2017). Nevertheless, the major storm of 1872 is a salutary reminder that intense storms such as the Carrington storm may occur more frequently than originally thought.

Among these extreme magnetic storms, Chapman (1957), and Silverman and Cliver (2001) suggested that the major storm in 1872 rivals the Carrington storm in 1859. Cliver & Svalgaard (2004) placed this event in the top rankings of magnetic disturbance at Greenwich, the *aa* index and the equatorward extension of auroral visibility. Willis et al. (2007) surveyed auroral records in East Asia after the 1840s and compared them with magnetic observations. These authors suggested that the large magnetic storm on February 4, 1872 lasted until February 6. Silverman (2008) investigated the records of auroral observations on the same date from all over the world, and also evaluated the equatorward extension of the auroral visibility in terms of dipole coordinates of the International Geomagnetic Reference Field (IGRF) in 1900. His research confirmed that this particular auroral display is comparable to the auroral display during the Carrington storm of 1859, in terms of its equatorward extension of visibility. As such, it can be regarded as the most equatorward aurora since the 1859 event. Silverman (2008) confirmed that this event at least competes with the Carrington event in terms of the equatorward extent of auroral visibility.

Therefore, we re-examined East Asian historical documents for this major storm, particularly records of naked-eye sunspot observations and auroral observations including an auroral drawing from Japan. We attempted to reconstruct the equatorward extension of the auroral visibility and the auroral boundary in terms of contemporary dipole magnetic coordinates. We also consider the duration of magnetic disturbance, the color and brightness of the aurora in terms of the flux of high-intensity low-energy electrons, and contemporary sunspot activity — all in the context of analyzing this major magnetic storm.

## 2. Method and Source Documents

We examined contemporary historical documents in East Asia for records of sunspots and auroral displays in February 1872, since East Asia was situated at relatively low magnetic latitude due to the location of the geomagnetic pole at that time. We surveyed the official histories and local treatises in China, diaries in Korea, in addition to diaries and chronicles in Japan, during the month of February 1872. In Japan, we obtained some diaries, chronicles, memoires, and reports of contemporary interviews with intellectuals and scholars of the time. The source documents with records of sunspots or auroral displays are shown in the Appendix 1.1. We summarized the observational records in terms of their date, color and observational time in Table 1.

We then considered the relevant observational sites. The official historical documents and local documents provided the observational details of where they were written and compiled, or otherwise endorsed (e.g. Beijing Observatory, 1985; Stephenson, 1997). Geographical coordinates of each observational site are provided in Table 1. Based on this information, we computed the magnetic latitude (hereafter, MLAT) of the observational sites, since it is known that the magnitude of magnetic storms is correlated with the equatorward boundaries of the auroral oval (e.g. Yokoyama et al., 1998). The magnetic north pole was computed for 1872 based on the historical magnetic field model GUFM 1 (Jackson et al., 2000).

For the sunspot record, we contextualize it in terms of longer-term solar activity. With regard to the auroral activity, the observational records were analyzed in terms of their equatorward extension, their color and brightness, and their duration in comparison with simultaneous magnetic observations.

## 3. Results

Within the contemporary historical documents archived in East Asia, we identified 1 sunspot record from China and 48 auroral records (7 from China, 4 from Korea, and 37 from Japan). The observational dates were converted to the luni-solar calendar and then to the Gregorian calendar based on the work of Wang et al. (2006) for Chinese dates and Uchida (1993) for Japanese dates. We also found one record of a naked-eye sunspot observation in a Chinese local treatise. The auroral records referred to the nights of February 4 to February 6. We provide their observational date, color, direction, observational time, geographical coordinate, and MLAT, together with their ID. Each ID refers to a historical document in the section for historical sources in the Supplementary Material (Table 1). The auroras were mostly red in color and typically appeared in a northerly direction although they sometimes extended beyond the zenith (C1, C3, J11, and J12). The observational sites roughly spanned 30° to 40° in geographical latitude, and were computed to span from 18.7° to 30.3° in MLAT.

## 4. Equatorward Extension of Auroral Display

It is known that the equatorward-most boundary of the auroral oval correlates with the magnitude of the minimum Dst value of geomagnetic storms (e.g. Yokoyama et al., 1998). The magnetic latitude of each observational site in East Asia was calculated based on their geographical coordinates according to the historical magnetic field model GUFM1 (Jackson et al., 2000). Note that Willis et al. (2007) computed the magnetic latitude using the corrected geomagnetic coordinates of the DGRF/IGRF model for the year 1900 and Silverman (2008) computed it using the dipole coordinates of the same model in the year 1900. Accordingly, we computed the location of the magnetic north pole, for the dipole component of the geomagnetic field, as N 78.57°, E 292.83°. The MLAT of each observational site was then calculated based on the distance from the magnetic north pole as shown in the section of the MLAT in Table 1.

Table 1 and Figure 1 show that the visibility of the auroral display in East Asia spanned the range 18.7° to 30.3° in MLAT. The most equatorward observational site based on MLAT is Shàoxīng (C6: N30°00′, E120°35′; 18.7° MLAT) on February 4, followed by Shanghai (C5: N31°14′ E121°29′; 19.9° MLAT) during the same night. The Italian consulate in Shanghai reported auroral display in the zenith: "the bright arc extended almost for 50 degrees very close to the Zenith". If this is correct, we can estimate the equatorward boundary of the auroral display as 24.2° invariant latitude (ILAT) on February 4. The ILAT has been used to identify the magnetic field line. ILAT is constant along a field line, whereas MLAT is dependent on the latitude (or the radial distance). On the ground, ILAT is equal to MLAT. We computed the ILAT according to the procedure of O'Brien et al. (1962), estimating the altitude of aurora as 400km as in Ebihara et al. (2017).

This description and our resultant estimation are supported by the information related to the direction of the auroral displays, since some records indicate that auroral displays were seen not only in the northern sky but also in other regions of the sky. Chinese records show the auroral display extended even beyond the zenith "from northeast to southwest" on February 4, 1872 at Dōngguāng-xiàn (C1: N37°53′, E116°32′; 26.5° MLAT) and on February 5, 1872 at Běijīng-xiàn (C3: N37°41′, E116°16′; 26.3° MLAT). Japanese records indicate that auroral display illuminated the entire sky around Kosugaya (J11: N34°50′, E136°52′; 24.3° MLAT) and extended "from north to south" at Hamada (J32: N34°54′, E132°05′; 24.0° MLAT) on February 4. More importantly, we found that an auroral drawing was produced by Buddhists at Okazaki (J9: N34°57′, E137°10′; 24.4° MLAT) as reproduced in Figure 2. This drawing has a caption stating "three bands of red vapor appreared in the western sky. Some rumored Nagoya was in conflagration, and others in Owari and Mino rumored Kyoto was in conflagration". This record explicitly shows that the western sky was red and fiery with considerable brightness. In contrast, this drawing (Figure 2) shows a significantly inclined auroral display from north to east. We need to consider the reliability of this drawing, which was made according to hearsay. This drawing is included in *Junkyo Eshi* (*Graphical History of Martydom*) compiled by Tanaka Nagane (1849–1922), who came to the Shounji Temple and made the drawing in question by interviewing local people near Okazaki and Nishio, and completed the drawing in 1911. Therefore, it is thought that Tanaka Nagane painted this drawing according to hearsay and mistakenly reversed the direction of the auroral inclination.

Therefore, we consider that the equatorward auroral boundary should extend down to 24.2° ILAT (19.9° MLAT) and 29.6° ILAT (26.3° MLAT) in their maximum phases on February 4 and 5, 1872. The great auroral display was also seen on February 6 as well. Yamanaka Masao, a principal of a high school in Hiroshima, witnessed both of the auroral displays on February 6, 1872 and on September 25, 1909. He compared both of them and considered "this [on February 6, 1872] was much larger than the recent one [on September 25, 1909]". During this time, he was off the coast of Enshu (J8: N34°43′, E137°44′; 24.2° MLAT) in the ship "Keiweru" from Kobe. Silverman (1995) investigated the great auroral display of September 25, 1909 and concluded that the auroral display was visible at least down to 30° MLAT. The description of J9 therefore confirms that the equatorward extension of the auroral display on February 6, 1872, was also much more intense than that of September 1909 and hence is consistent with the results of recent studies (Silverman, 1995, 2008; Cliver and Svalgaard, 2004; Cliver and Dietrich, 2013).

Assuming the equatorward-most auroral boundary, we attempted to scale the magnetic storms in term of their Dst values based on the equatorward extension of the auroral displays. By applying the formulae of Yokoyama et al. (1998) for 24.2° ILAT (C5) of the overhead aurora, we estimate their Dst value as approximately −1900 nT. This value is comparable with the value obtained for the Carrington storm in September 1859 with its horizontal component at Colaba scaled −1600 nT (Tsurutani et al., 2003; Nevanlinna, 2008; Kumar et al., 2015). It must be noted that this estimation for the Dst value is no more than an estimation due to the ambiguity of the equatorward extrapolation of the contemporary auroral oval, and the deviation of the models in Yokoyama et al. (1998).

## 5. Color and Brightness of Auroral Displays

The colors of these auroral displays are mostly reported as reddish and hence considered as type d aurora (see, Vallance Jones, 1971). Notable exceptions are found in C3 (26.3° MLAT), C6 (18.7° MLAT), and K2 (26.5° MLAT) reported as "a star colored in indigo blue" which "disappeared after a while", "celestial conflagration in purple", and "purple vapor in the north". If these "purple" and "indigo" displays show bluish aurora, they might indicate auroral displays at wavelength 427.8 nm ($N_2^+$ first negative band) (Tinsley et al., 1984; Zhang et al., 2006).

The brightness of these auroral displays is frequently compared with conflagrations or large fires, as this description is peculiar to bright auroral displays in extreme space weather events. During the Carrington storm, a large number of observers in North America compared the brightness of the auroral display with conflagrations or large fires (e.g. Green et al., 2006; Odenwald, 2007; Hayakawa et al., 2016b).

An observer at Hamada (J4: N34°54′, E132°05′; 24.0° MLAT) recollected the brightness of the auroral display on February 4 as it "enabled us to count the trees in the mountains and forests". This is considered to fall in the brightness of class IV of the International Brightness Coefficient (hereafter, IBC) that "provides a total illumination (at the ground) equivalent to full moonlight" (Chamberlain, 1961, p.124) or "casts discernible shadows" (Keay, 1990, p.378). The brightness is probably comparable to that witnessed on September 17, 1770, and the extremely bright aurora was probably caused by precipitation of high-intensity low-energy electrons (Ebihara et al., 2017).

## 6. Duration of these Auroral Displays in Comparison with Magnetic Observations

The ground magnetic field was recorded at Colaba, Bombay (Mayaud, 1973) as reproduced in Figure 3. A sudden start occurred at 19:18 Local Mumbai Time (LMT=UT+4.85 hours) on February 4 as characterized by a sudden increase in the H-component of the magnetic field. After the sudden start, the H-component of the magnetic field started to decrease at ~20:00 LMT, and reached a minimum at ~22:30 LMT. The amplitude of the decrease in the H-component of the magnetic field from the pre-storm level is estimated to be about −830 nT. The minimum depression of the H-component of the magnetic field occurred in the midnight-evening sector. The H-component of the magnetic field started to recover at ~23:00 LMT. The aurora recorded in China and Japan approximately corresponds to the initial phase, main phase and the early recovery phase of the magnetic storm. This is consistent with modern observations of aurora in Japan (Shiokawa et al., 2005).

During the storm's main phase, the magnetic depression on the dusk side is known to be larger than on the dawn side (Cummings, 1966; Fukushima and Kamide, 1973), which is probably due to the asymmetric distribution of the energy density of the hot ions in the inner magnetosphere (a major carrier of the storm-time ring current) (Ebihara et al., 2002). The magnetic disturbances on the dawn side are expected to be lower than on the dusk side (Colaba). Since the Dst value is derived by averaging the H-component disturbances observed by 4 stations at low latitudes, the amplitude of the Dst variation is probably smaller than that of the magnetic disturbance at dusk. It is speculated that the corresponding minimum Dst is probably larger than −830 nT. This may indicate that the Dst value derived from the equatorward-most auroral boundary is significantly overestimated.

The amplitude of the sudden commencement (SC) that occurred at 19:18 is estimated to be ~70 nT by visual inspection. Araki (2014) surveyed the sudden commencements recorded at Colaba and Alibag from 1871 to 1967. According to Araki (2014), the amplitude of the SC is classified as being extreme in terms of occurrence frequency. The amplitude of the sudden commencement $\Delta H$ is related to the solar wind dynamic pressure as $\Delta H = C\Delta P^{1/2}$, where $\Delta P$ is the jump of the solar wind dynamic pressure, and $C = 15$ nT/nPa$^{0.5}$ (Araki, 2014). Substituting $\Delta H$ of 70 nT, we estimate the jump of the solar wind dynamic pressure to be 22 nPa. The solar wind dynamic pressure is given by $P = mnV^2$, where $m$ is the mass of the solar wind (1.16 times proton mass), $n$ is the density of the solar wind, and $V$ is the solar wind speed. In order to achieve the solar wind dynamic pressure of 22 nPa, the solar wind speed is estimated to be 3350, 1059, 335 km/s for solar wind densities of 1, 10, and 100 cm$^{-3}$, respectively.

## 7. Comparison with Auroral Records in Other Sectors

The global extent of auroral displays other than in East Asia is intriguing. Chapman (1957b) introduced an article in the Indian newspaper (The Times of India, 1872-02-06, p.2) in which he reported an auroral observation at Bombay (N18°56′, E072°50′; 10.0° MLAT) and Aden. The newspaper in question states "After sunset on Sunday, the Aurora was slightly visible, and constantly kept changing colour, becoming deep violet, when it was most intense -- about three o'clock on Monday morning. It was visible until sunrise on Monday". This auroral observation at Bombay has been considered to be too equatorward, even in comparison with other Indian observational sites (e.g. Silverman, 2008). However, the equatorward extension of the auroral display in the East Asian sector suggests that this is not impossible. If the observer (who is situated at 10.0° MLAT) saw an auroral display within the elevation angle of 10° or 15° from the poleward horizon, the equatorward boundary of the auroral at 400km altitude would be calculated as 26.0° ILAT (22.1° MLAT) or 19.7° ILAT (24.0° MLAT), respectively. Note that a dipole magnetic field line is assumed in this calculation. This is consistent with the equatorward extension of auroral display in the East Asian sector at 24.2° ILAT (19.9° MLAT), estimated from C5.

The observations in the southern hemisphere also help us to consider the equatorward auroral extension. Meldrum (1872) reports the aurora australis seen at Mauritius (S20°10′, E57°31′; −26.3°MLAT) during the night of February 4, 1872. At approximately 23:00–23:20, he reports "Almost from the extreme left to the extreme right, and from as low down as I could see up to meridional altitude of about 72°, the sky was furrowed with alternate white and dark bands, all of which, so far as I could judge, were parallel to each other and to the magnetic meridian" (Meldrum, 1872, p.392). If the same auroral height of 400km is assumed, the equatorward boundary of the auroral oval in the Indian Ocean sector is calculated to be at least −27.4° MLAT (−30.6° ILAT) at its maximum. Its absolute value (30.6° ILAT) is probably not too different from the value of the equatorward extension of the auroral oval in the East Asian sector. This difference may be partially attributed to the different time, and the different local time of these observations. The aurora appeared throughout the entire sky around Iwatsuki (J12) post-midnight, whereas the aurora appeared at the meridional altitude of about 72° at Mauritius pre-midnight. It is expected that the aurora extended equatorward most in the Japan sector near midnight, and that the aurora moved poleward in the Mauritius sector near midnight. Mauritius is located about 80° westward of Iwatsuki (J12). If the aurora was observed at Mauritius and Iwatsuki simultaneously, the latitudinal difference of the auroral oval could be attributed to the spatial difference of the auroral oval, which extended equatorward compared to earlier local times.

## 8. Contemporary Solar Activity

The major storm in February 1872 occurred in the declining phase of Solar Cycle 11 after its maximum in 1870 (Clette et al., 2014; Svalgaard and Schatten, 2016; Vaquero et al., 2016). This temporal evolution is also confirmed by the normalized sunspot area series between 1832 and 2008 as shown in Figure 1 of Carrasco et al. (2016). Figures 4a and 4b show the monthly and daily sunspot number obtained from SILSO (Clette et al., 2014) as function of date and location of this major storm in time. Figure 4 indicates that the sunspot number started to decline from its maximum in May 1870 with a small peak in February 1872.

   In fact, a Chinese local intellectual reports a naked-eye sunspot observation at *Guăngān* (N30°28′, E106°38′) on February 9, 1872. There have been no previously reported naked-eye sunspot records for these days (Yau et al, 1988; Willis et al., 1996; Vaquero and Vazquez, 2009). This sunspot record (C0: *Guăngān-zhōuzhì*, v.13, f.5a) is shown in Figure 5 with a transcription and translation included in the Appendix 2. The date of the event is converted to February 9, 1872, where a considerably large number of sunspots were visible, with a few major sunspot groups on the right part of the Sun image. This might allow us to speculate that a large flare event could have happened a few days before the storm, due to the high activity of those Sun regions, when the sunspot groups were still on the left side of the Sun image according to the Sun rotation. Additional studies on the evolution of sunspots during that period (i.e. from the end of January 1872 to the first half of February 1872) requires extra material from further European sources, which are not included here, since it is outside the scope of this paper.

As mentioned above, although this unaided–eye sunspot observation occurred after the storm of February 4–February 6, 1872, it might be still connected with the events being probably present on the solar disk before the storm. However, other sunspot groups could have disappeared before the above-mentioned ones were moving close to the Sun center. This second assumption can be supported by occidental reports, which confirmed that large sunspots were seen in early February of 1872. Meldrum (1872, p.392) reported "On Friday a chain of spots stretched over nearly the whole of the sun's disc, and a large group occupied another part of it. On Monday, the chain had disappeared". Considering that this record is given in the context of the contemporary auroral display on February 4, it might mean another extended active region appeared on Friday February 2 and disappeared on Monday February 5, before the large active region on February 9.

At the very least, it is well known that sunspot number is highly correlated with sunspot area in the same period (Hathaway et al., 2002) and hence naked-eye sunspot observations are reported more in the active phase of the sunspot cycle (Heath, 1994). Therefore, we can securely conclude that the solar activity in early February 1872 was considerably enhanced.

## 9. Conclusion

In this article, we have investigated significant solar activity and a severe magnetic storm which occurred in February 1872, as observed mainly in East Asia. The East Asian historical records identified a naked-eye sunspot observation during the active phase of the solar activity in February 1872. These East Asian historical documents revealed an auroral display that was visible down to Shàoxīng (18.7° MLAT), with its equatorward boundary at 24.2° ILAT (19.9° MLAT). The equatorward boundary of the auroral oval suggested that the auroral observation in Bombay (10.0° MLAT) was probably valid within an elevation angle of 14°. We therefore confirm and establish that this event rivaled or even possibly surpassed the Carrington storm in September 1859, not only in terms of equatorward extension of auroral visibility (Chapman, 1957; Cliver and Svalgaard, 2004; Silverman, 2008), but also in terms of the equatorward extension of the auroral oval itself reconstructed from the contemporary records with information of elevation angle. These auroral records also assisted in elucidating the extreme brightness caused by the unusual precipitation of high-intensity low-energy electrons in February 1872. Moreover, the auroral observations corresponded with the initial phase, main phase, and the early recovery phase of the magnetic storm recorded by magnetic observations in Bombay. The amplitude of the magnetic disturbance in Bombay was about 830 nT. Therefore, we can reasonably consider that the great solar-terrestrial storm in February 1872 was certainly one of the most extreme solar-terrestrial storms in terms of its equatorward auroral extension, indeed competing with or even possibly surpassing the Carrington storm in 1859.

The auroral activity in the East Asian sector shows us these solar-terrestrial storms are worth consideration in other sectors in terms of equatorward extension of auroral oval, comparison with magnetic records, and analyses of contemporary sunspot active regions. These topics should be discussed elsewhere, especially as the structure of sunspot regions and their formation processes are highly related to the productivity of flare eruptions (Zirin & Liggett 1987; Sammis et al. 2000; Toriumi et al. 2017; Toriumi & Takasao 2017).

**Acknowledgements:**

We thank Mr. Y. Izumi for providing us with permission to pursue this research and the reproduction of *Junkyo Eshi* (J9) stored in Shounji, Dr. D. T. Rossi for his advice on the interpretation of Donati's Italian text, Dr. B. Veenadhari for her valuable suggestion regarding the magnetic field data recorded at Colaba, Mr. Y. Watanabe and Dr. K. Iwahashi for considerable advice on Japanese historical documents, Mr. H. J. Jo for advices on the Korean historical documents, and Dr. J. Williams and librarians in Lexington Library for significant help in obtaining copies of Donati's report. We gratefully acknowledge the support of Kyoto University's Supporting Program for Interaction-based Initiative Team Studies "Integrated study on human in space" (PI: H. Isobe), the Interdisciplinary Research Idea contest 2014 held by the Center for the Promotion of Interdisciplinary Education and Research, the "UCHUGAKU" project of the Unit of Synergetic Studies for Space, the Exploratory and Mission Research Projects of the Research Institute for Sustainable Humanosphere (PI: H. Isobe) and SPIRITS 2017 (PI: Y. Kano) of Kyoto University. This work was also supported by a Grant-in-Aid from the Ministry of Education, Culture, Sports, Science and Technology of Japan, Grant Number JP18H01254 (PI: H. Isobe), JP15H05816 (PI: S. Yoden), JP15H03732 (PI: Y. Ebihara), JP16H03955 (PI: K. Shibata), and JP15H05815 (PI: Y. Miyoshi), JP16K17671 (PI: S. Toriumi), JP15H05814 (PI: K. Ichimoto), and a Grant-in-Aid for JSPS Research Fellows JP17J06954 (PI: H. Hayakawa).

**References**

Araki, T. 2014, Earth, Planets and Space, 66, 164.

Baker, D. N., et al. 2008, Severe space weather events—understanding societal and economic impacts. National Academies Press, Washington DC.

Baker, D. N., Li, X., Pulkkinen, A., Ngwira, C. M., Mays, M. L., Galvin, A. B., Simunac, K. D. C. 2013, Space Weather, 11, 10, 585-591. doi: 10.1002/swe.20097

Beijing Observatory, 1985, An integrated catalogue of Chinese local treatises (Beijing: Zhinghua Book Company) [in Chinese]

Boteler, D. H. 2006, Advances in Space Research, 38, 2, 159. doi: 10.1016/j.asr.2006.01.013

Carrasco, V. M. S., Vaquero, J. M., Gallego, M. C., Sánchez-Bajo, F. 2016, Solar Physics, 291, 9-10, 2931-2940. doi: 10.1007/s11207-016-0943-9

Carrington, R.C. 1859, MNRAS, 20, 13. doi: 10.1093/mnras/20.1.13

Chamberlain, J. W. 1961, Physics of the Aurora and Airglow, pp. 124-125.

Chapman, S. 1957, Nature, 179, 4549, 7. doi: 10.1038/179007a0

Clette, F., Svalgaard, L., Vaquero, J.M., Cliver, E.W. 2014, Space Sci. Rev. 186, 1, 35. doi: 10.1007/s11214-014-0074-2

Cliver, E.W., Dietrich, D.F. 2013, Journal of Space Weather and Space Climate, 3, A31. doi: 10.1051/swsc/2013053

Cliver, E.W., Svalgaard, L. 2004, Solar Physics, 224, 407.

Cummings, W. D. (1966), Asymmetric ring currents and the low-latitude disturbance daily variation, J. Geophys. Res., 71(19), 4495–4503, doi:10.1029/JZ071i019p04495.

Curto, J.J., Castell, J., Del Moral, F. 2016, Journal of Space Weather and Space Climate, 6, A23. doi: 10.1051/swsc/2016018

Ebihara, Y., M. Ejiri, H. Nilsson, I. Sandahl, A. Milillo, M. Grande, J. F. Fennell, and J. L. Roeder (2002), Statistical distribution of the storm-time proton ring current: POLAR measurements, Geophysical Research Letters, Vol.29, No.20, 1969, doi:10.1029/2002GL015430.

Ebihara. Y., Hayakawa, H., Iwahashi, K., Tamazawa, H., Kawamura, A.D., Isobe, H. 2017, Space Weather, 15, 1373. doi: 10.1002/2017SW001693

Fox, N. J., Velli, M. C., Bale, S. D., et al. 2016, Space Science Reviews, 204, 1-4, 7-48. doi: 10.1007/s11214-015-0211-6

Fukushima, N., and Y. Kamide (1973), Partial ring current models for worldwide geomagnetic

disturbances, Rev. Geophys., 11(4), 795–853, doi:10.1029/RG011i004p00795.

Green, J., Boardsen, S. 2006, Adv. Space Res., 38, 130

Green, J., Boardsen, S., Odenwald, S, Humble, J, Pazamickas, K.2006, Adv. Space Res., 38, 145

Hapgood, M. 2011, Advances in Space Research, 47, 12, 2059. doi: 10.1016/j.asr.2010.02.007

Hathaway, D. H., Wilson, R. M., Reichmann, E. J. 2002, *Solar Physics*, 211, 1, 357. doi: 10.1023/A:1022425402664

Hayakawa, H., Iwahashi, K., Ebihara, Y., et al. 2017c, The Astrophysical Journal Letters, 850, 2, L31, doi: 10.3847/2041-8213/aa9661

Hayakawa, H., Iwahashi, K., Tamazawa, H., et al. 2016b, PASJ, 68, 99. doi: 10.1093/pasj/psw097

Hayakawa, H., Mitsuma, Y., Fujiwara, Y., et al. 2017b, PASJ, 69, 17. doi: 10.1093/pasj/psw128

Hayakawa, H., Tamazawa, H., Uchuyama, Y., Ebihara, Y., Miyahara, H., Kosaka, S., Iwahashi, K., Isobe, H. 2017a, Sol. Phys., 292, 1, 12. doi: 10.1007/s11207-016-1039-2

Heath, A. W. 1994, *Journal of the British Astronomical Association*, 104, 6, 304.

Hodgson, R. 1859, Monthly Notices of the Royal Astronomical Society, 20, 15. doi: 10.1093/mnras/20.1.15

Jackson, A., Jonkers, A. R. T., Walker, M. R. 2000, Roy Soc of London Phil Tr A, 358, 1768, 957. doi: 10.1098/rsta.2000.0569

Kaiser, M. L., Kucera, T. A., Davila, J. M., St. Cyr, O. C., Guhathakurta, M., Christian, E. 2008, Space Science Reviews, 136, 1-4, 5-16. doi: 10.1007/s11214-007-9277-0

Keay, C. S. L. 1990, Journal of the Royal Astronomical Society of Canada, 84, 6, 373.

Kimball, D.S. 1960, A study of the aurora of 1859. Scientific Report No. 6 (Fairbanks: University of Alaska)

Kumar, S., Veenadhari, B., Tulasi Ram, S., Selvakumaran, R., Mukherjee, S., Singh, R., Kadam, B. D. 2015, Journal of Geophysical Research: Space Physics, 120, 9, 7307. doi: 10.1002/2015JA021661

Lakhina, G.S., Tsurutani, B.T. 2016, Geoscience Letters, 3, 5. doi: 10.1186/s40562-016-0037-4

Lefèvre, L., Vennerstrøm, S., Dumbović, M., et al. 2016, Solar Physics, 291, 5, 1483. doi: 10.1007/s11207-016-0892-3

Lockwood, M., Owens, M. J., Barnard, L., Scott, C. J., Usoskin, I. G. Nevanlinna, H. 2016, Solar Physics, 291, 9-10, 2811-2828. doi: 10.1007/s11207-016-0913-2

Lockwood, M., Owens, M., Barnard, L., Scott, C., Watt, C. Bentley, S. 2018, Journal of Space Weather and Space Climate, 8. A12. doi: 10.1051/swsc/2017048

Love, J. J. 2018, Space Weather, 16, 37–46. doi: 10.1002/2017SW001795

Love, J. J., Bedrosian, P. A., Schultz, A. 2017, Space Weather, 15, 658–662, doi:
        10.1002/2017SW001622.

Macdonald, L. T. 2015, Journal for the History of Astronomy, 46, 4, 469-490. doi:
        10.1177/0021828615609525

Macdonald, L. T. 2018, Kew Observatory and the Evolution of Victorian Science, 1840–1910,
        Pittsburg, University of Pittsburg Press.

Mayaud, P. N. 1973, IAGA Bulletin, 33

Mayaud, P. N. 1980, Derivation, Meaning, and Use of Geomagnetic Indices, AGU, Washington, D.
C..

Moreno Cárdenas, Cristancho Sánchez, S., Vargas Domínguez, S. 2016, Advances in Space
Research 57, 257-267.

Müller, D., Marsden, R. G., St. Cyr, O. C., Gilbert, H. R., The Solar Orbiter Team, 2013, Solar
Physics, 285, 1-2, 25-70. doi: 10.1007/s11207-012-0085-7

Nevanlinna, H. 2004, Ann. Geophys., 22, 1691-1704.

Nevanlinna, H. 2006, Advances in Space Research, 38, 180-187.

Nevanlinna, H. 2008, Advances in Space Research, 42, 1, 171. doi: 10.1016/j.asr.2008.01.002

Ngwira, C. M., Pulkkinen, A., Wilder, F. D., Crowley, G. 2013, Space Weather, 11, 3, 121-131. doi:
        10.1002/swe.20021

O'Brien, B. J., Laughlin, C. D., Van Allen, J. A., Frank, L. A. 1962, Journal of Geophysical
        Research, 67, 4, 1209-1225. doi: 10.1029/JZ067i004p01209

Odenwald, S. 2007, Space Weather, 5, S11005, doi: 10.1029/2007SW000344.

Odenwald, S. 2015, Solar Storms: 2000 Years of Human Calamity!, San Bernardini, Create Space.

Oughton, E., et al. 2016, Helios Solar Storm Scenario (Centre for Risk Studies, Univ. Cambridge)

Pulkkinen, A., Bernabeu, E., Eichner, J., Beggan, C., Thomson, A. W. P. 2012, Space Weather, 10, 4,
04003. doi: 10.1029/2011SW000750

Pulkkinen, A., Bernabeu, E., Thomson, A., et al. 2017, Space Weather, 15, 7, 828-856. doi:
10.1002/2016SW001501

Ribeiro, P., Vaquero, J. M., Trigo, R. 2011, Journal of Atmospheric and Solar-Terrestrial Physics, 73,
308-315.

Riley, P., Love, J. J. 2017, Space Weather, 15, 1, 53-64. doi: 10.1002/2016SW001470

Riley, P., Baker, D., Liu, Y.D., Verronen, P., Singer, H., Güdel, M. 2018, Space Science Reviews,

214, 1, 21. doi: 10.1007/s11214-017-0456-3

Rouillard, A. P., Davies, J. A., Forsyth, et al. 2009, Journal of Geophysical Research, 114, A7, A07106. doi: https://doi.org/10.1029/2008JA014034

Sammis, I., Tang, F., Zirin, H. 2000, The Astrophysical Journal, 540, 1, 583-587. doi: 10.1086/309303

Shiokawa, K., T. Ogawa, and Y. Kamide (2005), Low-latitude auroras observed in Japan: 1999–2004, J. Geophys. Res., 110, A05202, doi:10.1029/2004JA010706.

Silverman, S. M. 2008, Journal of Atmospheric and Solar-Terrestrial Physics, 70, 10, 1301. doi: 10.1016/j.jastp.2008.03.012

Silverman, S.M. 2006, Advances in Space Research, 38, 2, 136. doi: 10.1016/j.asr.2005.03.157

Silverman, S.M., Cliver, E.W. 2001, Journal of Atmospheric and Solar-Terrestrial Physics, 63, 5, 523. doi: 10.1016/S1364-6826(00)00174-7

Stephenson, F.R., 1997, Historical Eclipses and Earth's Rotation, Cambridge, UK: Cambridge University Press.

Svalgaard, L. Schatten, K.H. 2016, Solar Physics, 291, 9-10, 2653. doi: 10.1007/s11207-015-0815-8

Tinsley, B. A.; Rohrbaugh, R. P.; Rassoul, H., et al. 1984, Geophysical Research Letters, 11, 572. doi: 10.1029/GL011i006p00572

Toriumi, S., Schrijver, C.J., Harra, L.K., Hudson, H., Nagashima, K. 2017, The Astrophysical Journal, 834, 1, 56. doi: 10.3847/1538-4357/834/1/56

Toriumi, S., Takasao, S. 2017, The Astrophysical Journal, 850, 1, 39. doi: 10.3847/1538-4357/aa95c2

Tsurutani, B.T., Gonzalez, W.D., Lakhina, G.S., Alex, S. 2003, JGR, 108, A7. doi:10.1029/2002JA009504

Usoskin, I. G. 2017, Living Reviews in Solar Physics, 14, 1, 3. doi: 10.1007/s41116-017-0006-9

Vallance Jones, A. (1971). Auroral Spectroscopy. Space Science Reviews, 11(6), 776–826.

Vaquero, J. M., Svalgaard, L., Carrasco, V. M. S., et al. 2016, Solar Physics, 291, 9-10, 3061. doi: 10.1007/s11207-016-0982-2

Vaquero, J. M., Valente, M. A., Trigo, R. M., Gallego, M. C. 2008, JGR, 113, A08230.

Vaquero, J.M., Vázquez, M. 2009, The Sun recorded through history. Springer, Berlin.

Vennerstrom, S., Lefevre, L., Dumbović, M., et al. 2016, Solar Physics, 291, 5, 1447. doi: 10.1007/s11207-016-0897-y

Viljanen, A., Myllys, M., Nevanlinna, H. 2014, Journal of Space Weather and Space Climate, 4, A11.

doi: 10.1051/swsc/2014008

Willis, D. M., Armstrong, G. M., Ault, C. E., Stephenson, F. R. 2005, Annales Geophysicae, 23, 3, 945. doi: 10.5194/angeo-23-945-2005

Willis, D. M., Stephenson, F. R. 2001, Annales Geophysicae, 19, 3, 289. doi: 10.5194/angeo-19-289-2001

Willis, D. M., Stephenson, F. R., Fang, H. 2007, Annales Geophysicae, 25, 2, 417. doi: 10.5194/angeo-25-417-2007

Willis, D. M., Stephenson, F. R., Singh, J. R. 1996a, QJRAS, 37, 733

Willis, D. M., Davda, V. N., Stephenson, F. R. 1996b, QJRAS, 37, 189-229.

Yau, K.K.C., Stephenson, F.R.: 1988, QJRAS, 29, 175.

Yokoyama, N., Kamide, Y., & Miyaoka, H. 1998, Annales Geophysicae, 16, 566

Zhang, Y., Paxton, L.J., Kozyra, J.U., Kil, H., Brandt, P.C. 2006, J. Geophys. Res., 111, A09307, doi:10.1029/2005JA011152.

Zirin, H., Liggett, A. 1987, Solar Physics, 113, 1-2, 267-281. doi: 10.1007/BF00147707

Appendix 1: References of historical documents cited in this paper.

Appendix 1.1: References of East Asian historical documents

C0: 光緒四川廣安州志, 人文研 史-XI-4-P-240, v.13, f.5a

C1: 光緒河北東光縣志, 人文研 史-XI-4-A-344, v.11, f.20b

C2: 光緒湖北光化縣志, 人文研 史-XI-4-N-344, v.8, f.9a

C3: 民国河北景縣新志, 人文研 史-XI-4-A-336, v.14, f.18b

C4: 趙中孚編, 翁同龢日記排印本, v.2, 1970, p.640

C5: Donati, G.-B. 1874, Sul modo con cui propagarono i fenomeni luminosi della grande aurora polare osservato nella note dal 4 al 5 febbraio 1872, *Memorie del R. Osservatorio ad, Arcetri*, v.1, p.8

C6: 越縵堂日記, 清朝四大日記之一 越縵堂日記, v.5, 1963, p.2909

C7: 峴樵山房日記, 清季洪洞董氏日記六種, v.3, 1997, p.303

K1: 響山日記, 韓国史料叢書第 31 響山日記, 1985, p.104

K2: 沈遠權日記, 韓国史料叢書第 48 沈遠權日記, v.1, 2004, p.67

K3: 羅巖隨録, 韓国史料叢書第 27 羅巖隨録, 1980, p.18

K4: 朴氏家 日記, 韓国学史料叢書第 31 朴氏家 日記, v.2, 2004, p.582

J1: 明五石見の震災（那賀郡鍋石村藤井宗雄老人の筆記摘要）：地学雑誌, v.1, 3, 1889, p.89

J2: 明五石見の震災（常時邇摩郡村大森花勘増田齢造の話）：地学雑誌, v.1, 3, 1889,p.89

J3: 明五石見の震災（第二回）（那賀郡浅利村島田氏及び大家本郷村某）：地学雑誌, 1, 4, 1889, p.140.

J4: 明治五年ノ濱田地震, 震災豫防調査會報告, v.77, 1913, p.74

J5: 桜斎随筆 第十三冊, 桜斎随筆, v.4, 2000, p.111

J6: 年暦算, 遠賀郡鬼津村井ノ口家年暦算：幕末二百年間の遠賀郡郷土史総まくりの年代記, 1985, p.191

J7: 依岡宇兵衛諸事控, 印南町史 史料編, 1987, pp.1022-1024

J8: 気象集誌, v.11, 1909, p.406; 中国新聞 1909-09-28, p.4

J9: 奇瑞, 明治辛未殉教絵史, 1911, p.32

J10: 遠藤家々督自知録, 出雲市誌, 1951, p.896

J11: 日乗, 溝口 幹「日乗」v.4, 2010, p.69

J12: 万代記録帳, 岩槻市史, v.4, 1982, p.1044

J13: 廻瀾始末, 明治仏教全集, v.8, 1935, p.51

J14: 蓮専寺記録四, 由良町誌 史(資)料編, 1985, p.857

J15: 見聞略記巻之拾, 見聞略記 幕末筑前浦商人の記録, 1989, p.624

J16: 玉村清斎晴雨日記, いわき史料集成, v.4, 1990, p.288

J17: 歳旬漫筆, 日本庶民生活史料集成, v.12, 1971,p.240

J18: 工藤賜日記抄, 鹿角市史資料編, v.1, 1979, p.90

J19: 吉光寺日記, 小山市史, v.20, 1981, p.817

J20: 壬申随筆, 小泉蒼軒日録, v.2, 1974, pp.747-748

J21: 多志南美草, 多志南美草, v.4, 2006, pp.450-451

J22: 諸色書留帳, 大館市史編さん調査資料, v.16, 1975, p.65

J23: 藤井此蔵一生記, 日本庶民生活史料集, v.2, 1969, p.807, 809

J24: 明治四年 日記 辛未正月, 白根市史, v.4, 1986, p.483

J25: 御用日記, 比内町史資料, v.10, 1996, p.300

J26: 富田高慶日記,富田高慶日記, 1981, p.769

J27: 明治五歳正月 都鄙新聞第二号, 日本初期新聞全集, v.34, 1992, p.17

J28: 明治壬申正月 名古屋新聞第三号, 日本初期新聞全集, v.34, 1992, p.181

J29: 中原嘉左右日記, 中原嘉左右日記, v.2, 1970, p.514

J30：岡田親之日記, 幕末維新期の胎動と展開 : 栃木の在村記録（栃木市史料叢書 ；第 1 集）,v.3, 2016, p.645

J31: 読書室年表, 幕末本草家交信録 : 畔田翠山・山本沈三郎文書, 1996, p.257

J32: 日次記, 日本歴史 1996 年 3 月号, 明治五年石見浜田大地震の記録, 1996, p.69

J33: 明石家記録, 福島市史資料叢書. v.78, 2002, p.96,

J34: 暴動以後地震以前濱田縣の出来事, 浜田町史, 1935, p.283

J35: 浜田震災の話, 玉置啓太, 浜田町史, 1935, p.293

J36: 明治四辛未日記 野中氏, 棚倉町史別巻 2, 1982, p.164

J37: 明治四年辛未正月 日記 第十六号 Meiji 4-12-26 [秋田県立公文書館 児玉 20]

Appendix 1.2: References of occidental observations

Meldrum: Meldrum, J. 1872, Aurora Australis, *Nature*, v.5, pp.392-393.

Appendix 2: Transcription and translation of the naked-eye sunspot record in C0.

Transcription: 同治…十一年正月初日烏出其光如環中皆黒色遂隠

Translation: On the first of the first month of 11[th] year (Feb. 9, 1872), a crow was in the sun. Its light was as if everything in the ring was black. It finally disappeared.

Appendix 3: Transcription and translation of the caption of auroral drawing (J9; Figure 2)

Transcription: 十二月二十六日の夜半であった，西の空に三筋の赤氣が立つた，或は名古屋が大火だといひ，尾張や美濃では京都が火だと沙汰した。宛も此日太政官から囚徒の罪状が到達したので，或は其の徴だと後で噂された。

Translation: At 24:00 of 12[th] month 26[th] day (Feb. 4, 1872), three bands of red vapor appeared in the western sky. Some rumored Nagoya was in conflagration, and others in Owari and Mino rumored Kyoto was in conflagration. On this date, we received a decision letter of culpability from the Grand Council. Some rumored this phenomenon was omen of this event.

Table 1

| ID | Y | M | D | Color | Dir. | Start | End | Lat. | Long. | MLAT |
|----|------|---|---|-------|-------|-------|--------|---------|----------|------|
| C1 | 1872 | 2 | 4 | R | ne-sw | | dawn | N37°53′ | E116°32′ | 26.5 |
| C2 | 1872 | 2 | 4 | R | nw | 24:00 | sunrise | N32°24′ | E111°41′ | 21.0 |
| C3 | 1872 | 2 | 5 | R, B | ne-sw | 26:00 | | N37°41′ | E116°16′ | 26.3 |
| C4 | 1872 | 2 | 4 | R | | 26:00 | | N34°45′ | E113°41′ | 23.3 |
| C4 | 1872 | 2 | 4 | R | | 26:00 | | N34°16′ | E108°57′ | 22.9 |
| C4 | 1872 | 2 | 4 | R | | 26:00 | | N39°54′ | E116°26′ | 28.5 |
| C5 | 1872 | 2 | 4 | | n-z | | 26:30 | N31°14′ | E121°29′ | 19.9 |
| C5 | 1872 | 2 | 4 | R | | 24:30 | 27:15 | N39°07′ | E117°12′ | 27.7 |
| C6 | 1872 | 2 | 4 | Pu | n | 24:00 | | N30°00′ | E120°35′ | 18.7 |
| C7 | 1872 | 2 | 4 | | nw | 26:00 | 28:00 | N36°15′ | E111°41′ | 24.8 |
| K1 | 1872 | 2 | 4 | R | nw-ne | 24:00 | sunrise | N36°39′ | E128°27′ | 25.6 |
| K2 | 1872 | 2 | 4 | Pu | n | 28:00 | | N37°35′ | E126°59′ | 26.5 |
| K3 | 1872 | 2 | 4 | R/Bk | nw-ne | 22:00 | 27:00 | N35°32′ | E129°20′ | 24.5 |
| K4 | 1872 | 2 | 5 | R | nw-ne | | dawn | N36°39′ | E128°27′ | 25.6 |
| J1 | 1872 | 2 | 4 | R | n | | | N34°54′ | E132°05′ | 24.0 |
| J2 | 1872 | 2 | 6 | R | e-en | 26:00 | sunrise | N35°09′ | E132°24′ | 24.3 |
| J3 | 1872 | 2 | 6 | R | n | | | N34°40′ | E131°51′ | 23.8 |
| J4 | 1872 | 2 | 4 | R | nw | | | N34°54′ | E132°05′ | 24.0 |
| J5 | 1872 | 2 | 6 | R | nw-n | 28:00 | 30:00 | N36°22′ | E140°28′ | 26.1 |
| J6 | 1872 | 2 | 4 | R | n | 26:00 | dawn | N33°52′ | E130°39′ | 22.9 |
| J7 | 1872 | 2 | | R | ne-nw | | dawn | N33°59′ | E135°13′ | 23.3 |
| J8 | 1872 | 2 | 6 | R | n | | | N34°43′ | E137°44′ | 24.2 |
| J9 | 1872 | 2 | 4 | R | w | 24:00 | | N34°57′ | E137°10′ | 24.4 |
| J10 | 1872 | 2 | 4 | R | n-e | 27:00 | | N35°22′ | E132°45′ | 24.5 |
| J11 | 1872 | 2 | 4 | R | full | | | N34°50′ | E136°52′ | 24.3 |
| J12 | 1872 | 2 | 4 | R | nw-ne | 24:00 | dawn | N35°57′ | E139°42′ | 25.6 |
| J13 | 1872 | 2 | 4 | R | w | 26:00 | | N34°57′ | E137°10′ | 24.4 |
| J14 | 1872 | 2 | 4 | | n | | | N33°58′ | E135°07′ | 23.3 |
| J15 | 1872 | 2 | 4 | R | nw-ne | 22:00 | | N33°38′ | E130°14′ | 22.7 |
| J16 | 1872 | 2 | 4 | | nw | 26:00 | | N37°03′ | E140°53′ | 26.8 |
| J17 | 1872 | 2 | 4 | | n-w | | | N37°38′ | E138°53′ | 27.2 |
| J18 | 1872 | 2 | 4 | R | n | | | N40°13′ | E140°47′ | 29.9 |
| J19 | 1872 | 2 | 4 | | | | | N36°19′ | E139°48′ | 26.0 |

| J20 | 1872 | 2 | 4 | R | n-w | 22:00 | | N37°46′ | E139°01′ | 27.4 |
| J21 | 1872 | 2 | 5 | R | n | | | N40°31′ | E141°31′ | 30.3 |
| J22 | 1872 | 2 | 4 | R | nw | 28:00 | | N40°18′ | E140°35′ | 30.0 |
| J23 | 1872 | 2 | 4 | R | nw-ne | 22:00 | dawn | N34°16′ | E133°02′ | 23.5 |
| J24 | 1872 | 2 | 4 | | ne-nw | 24:00 | 30:00 | N37°41′ | E138°59′ | 27.3 |
| J25 | 1872 | 2 | 4 | R | e-w | 26:00 | | N40°14′ | E140°14′ | 29.9 |
| J26 | 1872 | 2 | 5 | R | n | | | N37°38′ | E140°56′ | 27.4 |
| J27 | 1872 | 2 | 4 | R | n | 26:00 | | N35°41′ | E139°45′ | 25.4 |
| J28 | 1872 | 2 | 4 | R | n | 24:00 | dawn | N35°11′ | E136°54′ | 24.6 |
| J29 | 1872 | 2 | 4 | R | e | 26:00 | | N33°53′ | E130°52′ | 23.0 |
| J30 | 1872 | 2 | 4 | R | n | | | N36°23′ | E139°44′ | 26.1 |
| J31 | 1872 | 2 | 4 | R | n | 27:00 | 27:30 | N35°00′ | E135°45′ | 24.4 |
| J32 | 1872 | 2 | 4 | R | n-s | 21:00 | 30:00 | N34°54′ | E132°05′ | 24.0 |
| J33 | 1872 | 2 | 4 | R | nw | 28:00 | | N37°29′ | E139°56′ | 27.2 |
| J34 | 1872 | 2 | 4 | | n | | | N34°54′ | E132°05′ | 24.0 |
| J35 | 1872 | 2 | 4 | | n | 24:00 | | N34°54′ | E132°05′ | 24.0 |
| J36 | 1872 | 2 | 4 | R | nw-ne | 26:00 | | N37°02′ | E140°23′ | 26.8 |
| J37 | 1872 | 2 | 4 | R | | 26:00 | | N40°06′ | E140°00′ | 29.8 |

Table 1: Catalog of observational reports from East Asia. Their data are given with their record ID, date (year, month, and date), color, direction, time of start and end, geographical latitude and longitude, and magnetic latitude (MLAT). The original reference is shown in Appendix 1 for each record ID. Their direction is given as the 8 points of the compass, unless otherwise endorsed. Their color is given as R (red), B (blue), Pu (purple), and Bk (black). The observational time is given as local time between 06:00 and 30:00 (06:00 on the following day). This system describe the observational time beyond midnight continuously with local time +24h on the same date (e.g., 25:00 on February 4 for 01:00 on February 5).

Figure 1

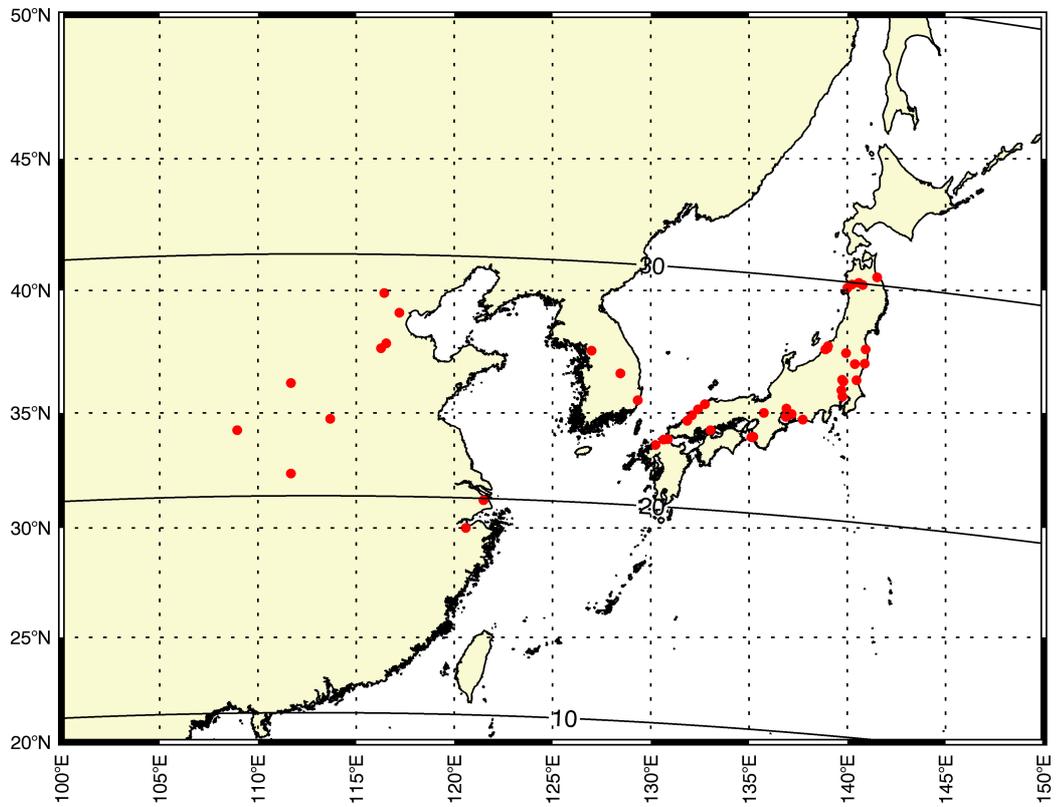

Figure 1: Observational sites of auroral displays during February 4-6, 1872, in East Asia. The broken lines show geographic latitude and longitude. The curved continuous lines show magnetic latitude.

Figure 2

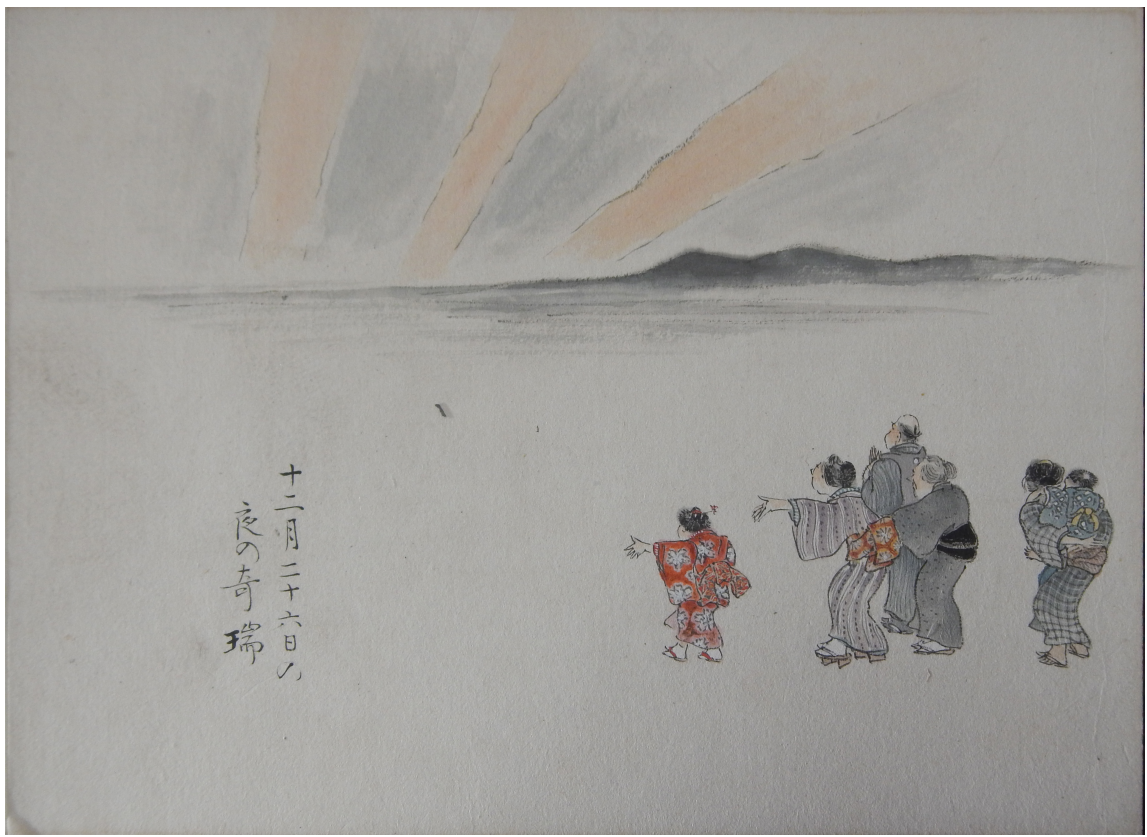

Figure 2: Auroral drawing of Junkyo Eshi (J9) representing auroral display on February 4, 1872. The original figure is currently preserved in Shounji Temple. This figure is reproduced with permission of Shounji Temple.

Figure 3

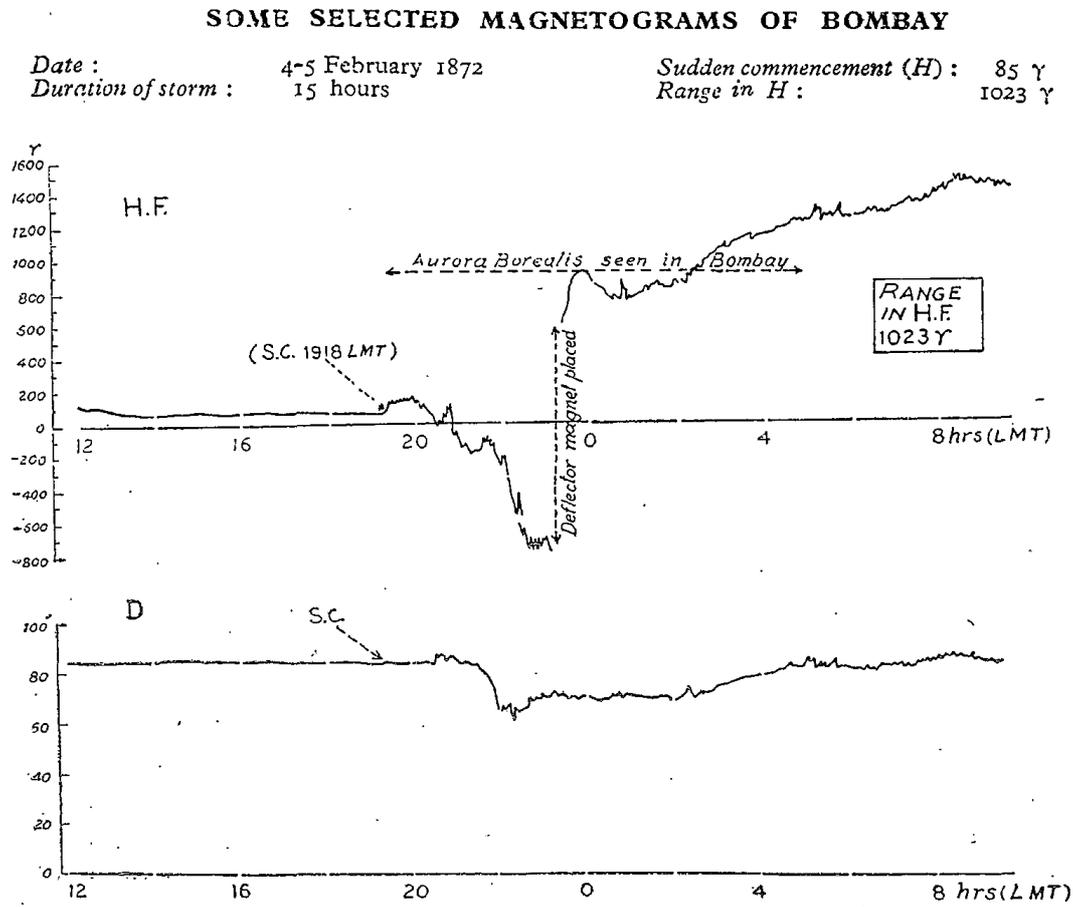

## SOME SELECTED MAGNETOGRAMS OF BOMBAY

*Date :*          4-5 February 1872          *Sudden commencement (H) :*   85 γ
*Duration of storm :*   15 hours          *Range in H :*                1023 γ

A magnetic storm of great intensity was recorded by Colaba on this day. It was one of the most violent storms since 1857 listed by Chapman and Bartels (*Geomagnetism* Vol. I page 328).

Very intense Aurora Borealis was seen in many places and even in low latitudes like Bombay from the evening of 4th to the morning of 5th. The Aurora Borealis of this date is one of the most brilliant on record. Internal telegraphic Communication and cable communication with England were interrupted for some hours as a result of the storm.

(Bombay LMT is 4 hours 51 minutes ahead of G M T)

Figure 3: Magnetic observation at Bombay during 4-5 Febraury, 1872, adopted from Mayaud (1973).

Figure 4

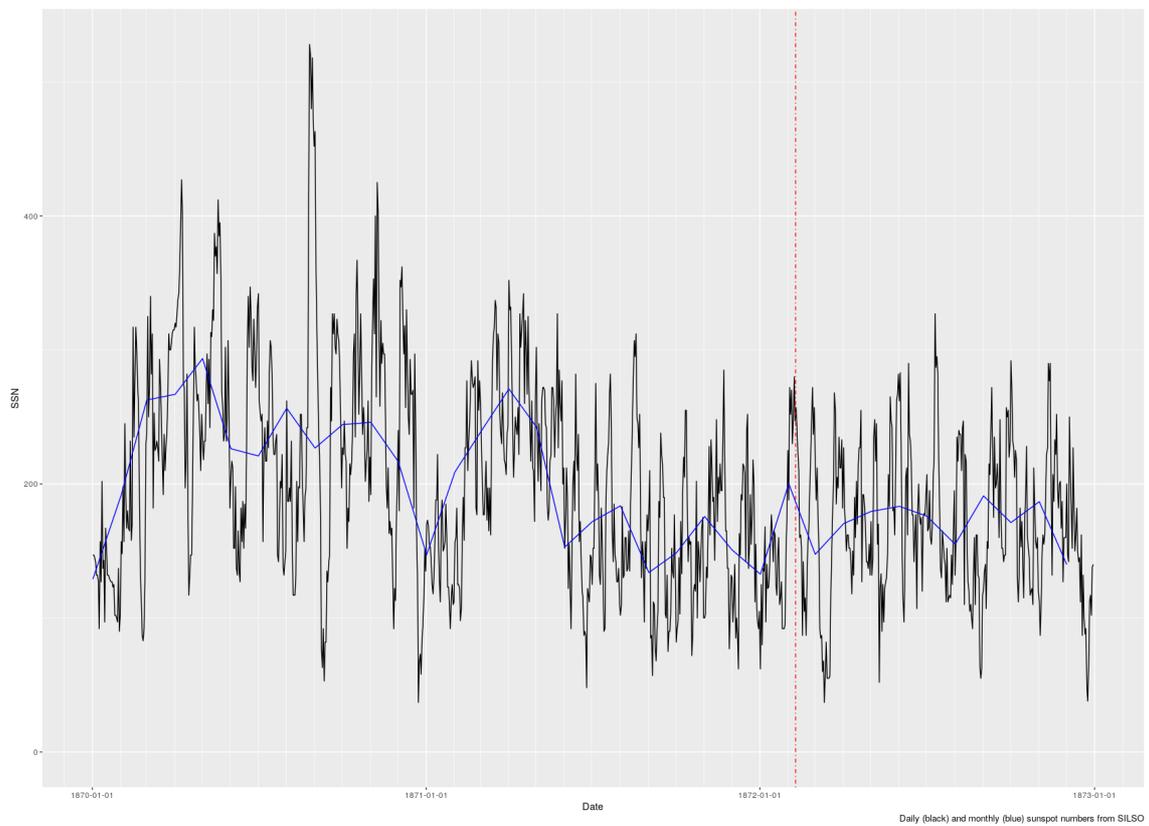

Figure 4: Monthly sunspot number (blue curve) and daily sunspot number (black curve) between 1870 and 1873 obtained from SILSO, as a context of the major storm in 1872 (a red broken stroke).

Figure 5

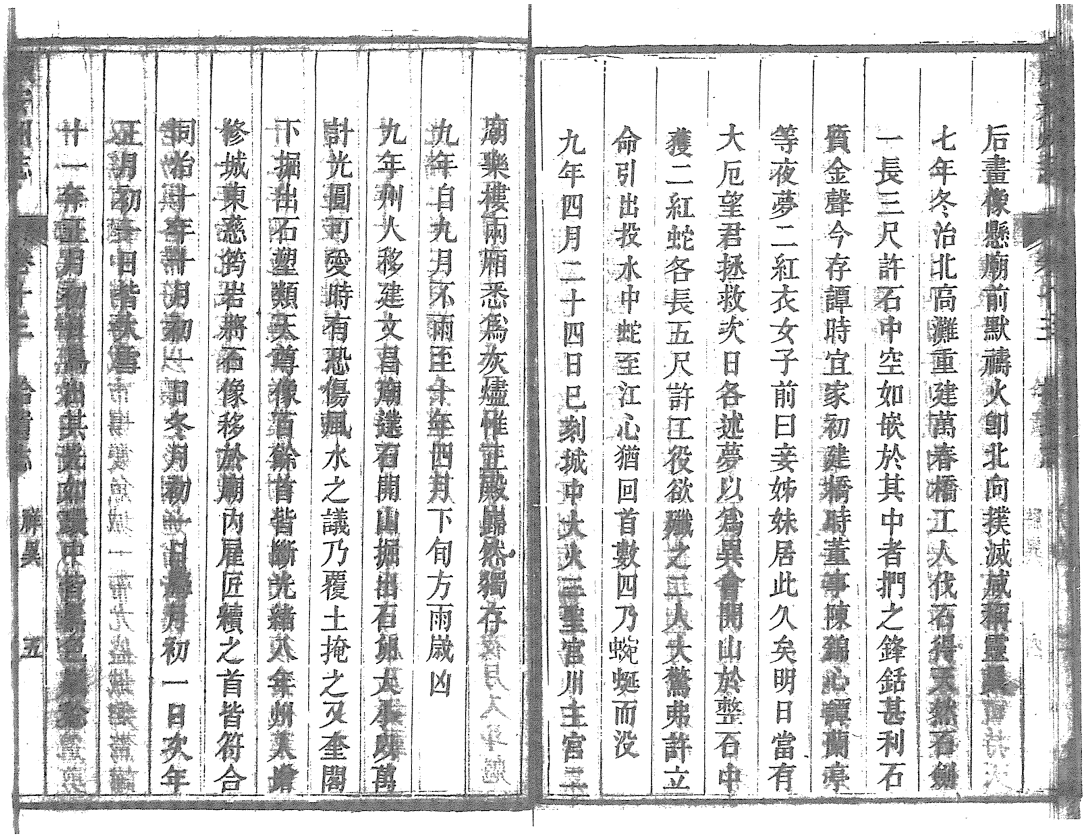

Figure 5: A report of naked-eye sunspot observation at *Guǎngān* (N30°28′, E106°38′) on February 9obtained from *Guǎngān-zhōuzhì* (v.13, f.5a). This figure is reprodiuced with permission of Institute for Research on Humanities of Kyoto University.